\documentclass[aps,prb,twocolumn,showpacs]{revtex4-1}
\usepackage[T1]{fontenc}
\usepackage[ascii]{inputenc}
\usepackage{url}
\usepackage{latexsym}
\usepackage{amssymb}
\usepackage{graphicx}
\usepackage{amsmath}
\usepackage{bm}
\usepackage[colorlinks=true,urlcolor=blue,linkcolor=blue,citecolor=blue]{hyperref}

\makeatletter

\newcommand*\MY@leftharpoonupfill@{%
    \arrowfill@\leftharpoonup\relbar\relbar
}
\newcommand*\MY@rightharpoonupfill@{%
    \arrowfill@\relbar\relbar\rightharpoonup
}
\newcommand*\overleftharpoon{%
    \mathpalette{\overarrow@\MY@leftharpoonupfill@}%
}
\newcommand*\overrightharpoon{%
    \mathpalette{\overarrow@\MY@rightharpoonupfill@}%
}

\makeatother

\begin{document}
\title{Effects of electron-impurity scattering on density of states in silicene: impurity bands and band-gap narrowing}
\author{S. Y. Liu}
\email{liusy@sjtu.edu.cn}
\affiliation{Key Laboratory of Artificial Structures and Quantum Control (Ministry of Education), Department of Physics and Astronomy, Shanghai
  Jiao Tong University, 800 Dongchuan Road,
  Shanghai 200240, China}
\affiliation{Collaborative Innovation Center of Advanced Microstructures, Nanjing 210093, China}
\author{Y. C. Zeng}
\affiliation{Key Laboratory of Artificial Structures and Quantum Control (Ministry of Education), Department of Physics and Astronomy, Shanghai
  Jiao Tong University, 800 Dongchuan Road,
  Shanghai 200240, China}
\affiliation{Collaborative Innovation Center of Advanced Microstructures, Nanjing 210093, China}
\author{X. L. Lei}
\affiliation{Key Laboratory of Artificial Structures and Quantum Control (Ministry of Education), Department of Physics and Astronomy, Shanghai
  Jiao Tong University, 800 Dongchuan Road,
  Shanghai 200240, China}
\affiliation{Collaborative Innovation Center of Advanced Microstructures, Nanjing 210093, China}

\begin{abstract}

  Considering the interband correlation, we present a generalized multiple-scattering approach of Green's function to investigate the effects of electron-impurity scattering on the density of states in silicene.  The reduction of energy gaps in the case of relatively high chemical potential and the transformation of split-off impurity bands into band tails for low chemical potential are found. The dependency of optical conductivity on the impurity concentration is also discussed for frequency within the terahertz regime.

\end{abstract}


\maketitle

\section{ Introduction}

Recently, silicene, a single layer of silicon atoms, has attracted a great deal of experimental and
theoretical interest.\cite{PhysRevB.76.075131,PhysRevLett.102.236804,PhysRevB.79.115409,APL.96.261905,APL.97.223109,APL.98.081909,PhysRevLett.107.076802,PhysRevB.84.195430,PhysRevLett.108.155501,PhysRevLett.108.245501,doi:10.1021/nl203065e,PhysRevB.85.075423,PhysRevB.86.195405,PSSR:PSSR201206202,1367-2630-14-3-033003,1882-0786-5-4-045802,Kara20121,0953-8984-24-22-223001,doi:10.1021/nl301047g,PhysRevLett.109.056804,APL.102.043113,APL.102.162412,PhysRevB.87.235426,PhysRevB.88.245408,doi:10.1021/nl304347w,0953-8984-26-34-345303,aizawa2014silicene,ADMA:ADMA201304783,doi:10.1021/ar400180e,voon2014silicene,PhysRevB.91.115411,Tao2015,Dimoulas201568,0953-8984-27-25-253002,0953-8984-27-20-203201,2053-1583-3-1-012001,PSSR:PSSR201510338,zhao2016rise,spencer2016silicene}  This two-dimensional system has a hexagonal honeycomb structure, similar to the graphene, but with a periodically buckled topology. Due to the strong intrinsic spin-orbit coupling (SOC), the energy gaps of silicene near Dirac cones are relatively larger than those in graphene. The magnitude of the energy gap due to intrinsic SOC may reach the value about $2\Delta_{SO}=1.55\sim 7.9$\,meV ($\Delta_{SO}$ is the characteristic energy of this SOC).\cite{PhysRevLett.107.076802,PhysRevB.84.195430} Besides, the specific buckled structure enables us to control the energy gap of silicene by applying an external perpendicular electric field.\cite{doi:10.1021/nl203065e,PhysRevB.85.075423,1367-2630-14-3-033003,PhysRevB.88.245408} These properties make the silicene  a promising candidate for future electronic and spintronic applications.
In experiment, silicene has
been successfully fabricated via epitaxial growth on the Ag(111),\cite{APL.97.223109,PhysRevLett.108.155501,1882-0786-5-4-045802,doi:10.1021/nl301047g,PhysRevLett.109.056804} ZrB$_2$(0001),\cite{PhysRevLett.108.245501}  ZrC(111),\cite{aizawa2014silicene} Ir(111),\cite{doi:10.1021/nl304347w} and MoS$_2$
surfaces\cite{ADMA:ADMA201304783} and the silicene field effect transistor (FET) operating at
room temperature has also been realized very recently.\cite{Tao2015} In theory, many interesting phenomena in silicene, such as the phase transition from a quantum spin-Hall state to a trivial insulating state \cite{PhysRevLett.107.076802,PhysRevB.86.195405,APL.102.162412,PhysRevB.87.235426,PSSR:PSSR201206202,APL.102.043113,0953-8984-26-34-345303}, the intrinsic spin-Hall and valley-Hall effects induced by ac and dc electric field\cite{PhysRevB.87.235426,0953-8984-26-34-345303,PhysRevB.91.115411}, {\it etc}. have been predicted. 

The magnitudes of energy gaps in materials play key roles in electronic device designing and development. They are also essential for the observation of many fundamental effects in condensed matter physics, such as quantum spin-Hall effect, quantum anomalous Hall effect, {\it etc}. However, the previous studies in bulk semiconductors indicated that the impurities may strongly affect the energy gaps.\cite{shklovskii2013electronic} When the concentration of impurities is relatively dilute, electron-impurity scattering may introduce  discrete energy levels within energy gaps. However, as the impurity density increases, additional bands form within or/and out of the energy gaps. These impurity bands (IBs) are further transformed into the band tails in highly doped semiconductors, leading to the reduction of energy gaps. Similar phenomena were also shown in conventional two-dimensional electron gases:\cite{PhysRevB.37.4589,PSSB:PSSB375,apl-102-162101} when the density of impurities increases, the width of split-off impurity band increases and the IB is finally transformed into the band tail. The dependency of band tailing on the dopant concentration in heavily n-type doped superlattices and single wells was also studied.\cite{PhysRevB.30.3367,PhysRevB.39.3400} However, it still remains unclear the effects of electron-impurity scattering on energy gaps in novel two-dimensional (2D) systems such as in graphene, silicene, and germanene, {\it etc.} where the band gaps are relatively small and the scattering may lead to strong interband correlation, which plays a substantial role in the study of electronic states.

To investigate the impurity problems in conventional bulk and low-dimensional semiconductors, many theoretical approaches have been proposed, including coherent-potential approximation,\cite{economou2006green} path-integral approach,\cite{samathiyakanit1974path,PhysRevB.19.2266,PhysRevB.22.6222,PhysRevB.30.3367,PSSB:PSSB375,apl-102-162101} semiclassical models,\cite{PhysRev.148.722,PhysRevB.44.12822,RevModPhys.64.755} instanton method,\cite{PhysRevB.37.6963} multiple-scattering approach (MSA),\cite{PhysRevLett.48.886,PhysRevB.28.4704,PhysRevB.37.4589,PhysRevB.39.3400} {\it etc}. Among these, MSA, a Green's function (GF) method within multiple-scattering approximation, enables us to correctly describe the evolution of electronic band structure with the doped concentration. Hence, in present paper, we generalize the MSA in the presence of the interband coherence and present a theoretical study of the effects of electron-impurity scattering on the density of states (DOS) in silicene.  The dependencies of DOS on impurity density for various chemical potentials (CPs) are carried out. Further, we also investigate the optical conductivities versus frequency within terahertz regime.

The paper is organized as follows. In section II, we present the Green's function approach within the multiple-scattering approximation in the presence of interband correlation. The numerical results are shown in Section III.  Finally, we conclude our results in section IV and append the derivation of the multiple-band Kubo formula in Appendix.   

\section{Theoretical Formulation}
A two-dimensional massive Dirac fermion with momentum ${\bf k}\equiv (k_x,k_y)$ and electric charge $-e$ near the $K$ or $K'$ Dirac node in buckled silicene is described by a Hamiltonian of the form [$\lambda_{\eta\sigma}=\Delta_{z}-\eta\sigma\Delta_{SO}$]
\begin{eqnarray}
\check h_{\eta\sigma} ({\bf k})&=&v_F(\eta k_x\hat\tau_x+k_y\hat\tau_y)+\lambda_{\eta\sigma}\hat\tau_z,\label{H1}
\end{eqnarray}
where $\eta=\pm 1$ is the valley index for $K$ and $K'$, $\sigma=\pm 1$ is the spin index for spin up and down, $\hat\tau_i$ ($i=x,y,z$) represent the Pauli matrices, and $v_F\approx 5.0\times 10^{5}$\,m/s is the Fermi velocity of Dirac fermion in silicene. There exist two types of SOC in silicene: the intrinsic SOC with characteristic energy about  $\Delta_{SO}\approx 3.9$\,meV,\cite{PhysRevB.84.195430,1367-2630-14-3-033003} and the SOC induced by the hybridization of $p_z$ orbitals with $\sigma$ orbitals of silicon atoms. The last one, which is described by an energy $\Delta_z$, can be tuned by applying external electric field along $z$-direction.\cite{doi:10.1021/nl203065e} 

 Hamiltonian (\ref{H1}) in pseudo-spin basis can be diagonized: it reduces to a diagonal pseudo-helicity-basis Hamiltonian of the form $\hat h_{\eta\sigma}({\bf k})={\rm diag}[\varepsilon_{\eta\sigma,+}({k}),\varepsilon_{\eta\sigma,-}({k})]$ with $\varepsilon_{\eta\sigma\mu}({k})=\mu g_{\eta\sigma;k}$, $g_{\eta\sigma;k}\equiv \sqrt{v_F^2 k^2+\lambda_{\eta\sigma}^2}$, and $\mu=\pm$ as the helicity index. Correspondingly, the left- and right-helicity wave functions, $\Psi_{\eta\sigma\mu{\bf k}}({\bf r})$, take the forms $\Psi_{\eta\sigma\mu{\bf k}}({\bf r})=\psi_{\eta\sigma\mu}({\bf k}){\rm e}^{i{\bf k}\cdot{\bf r}}$ with $\psi_{\eta\sigma\mu}({\bf k})$ given by
\begin{equation}
\psi_{\eta\sigma\mu}({\bf k})=\frac{1}{\sqrt{2g_{\eta\sigma;k}(g_{\eta\sigma;k}-\mu\lambda_{\eta\sigma})}}\left (
\begin{array}{c}
\eta\mu v_Fk{\rm e}^{-i\eta\varphi_{\bf k}}\\
g_{\eta\sigma;k}-\mu\lambda_{\eta\sigma}
\end{array}
\right ).
\end{equation}
Here, $k$ and $\varphi_{\bf k}$ are the magnitude and angle of momentum ${\bf k}$, respectively.

\begin{figure}
   \centering
   \includegraphics[width=0.45\textwidth]{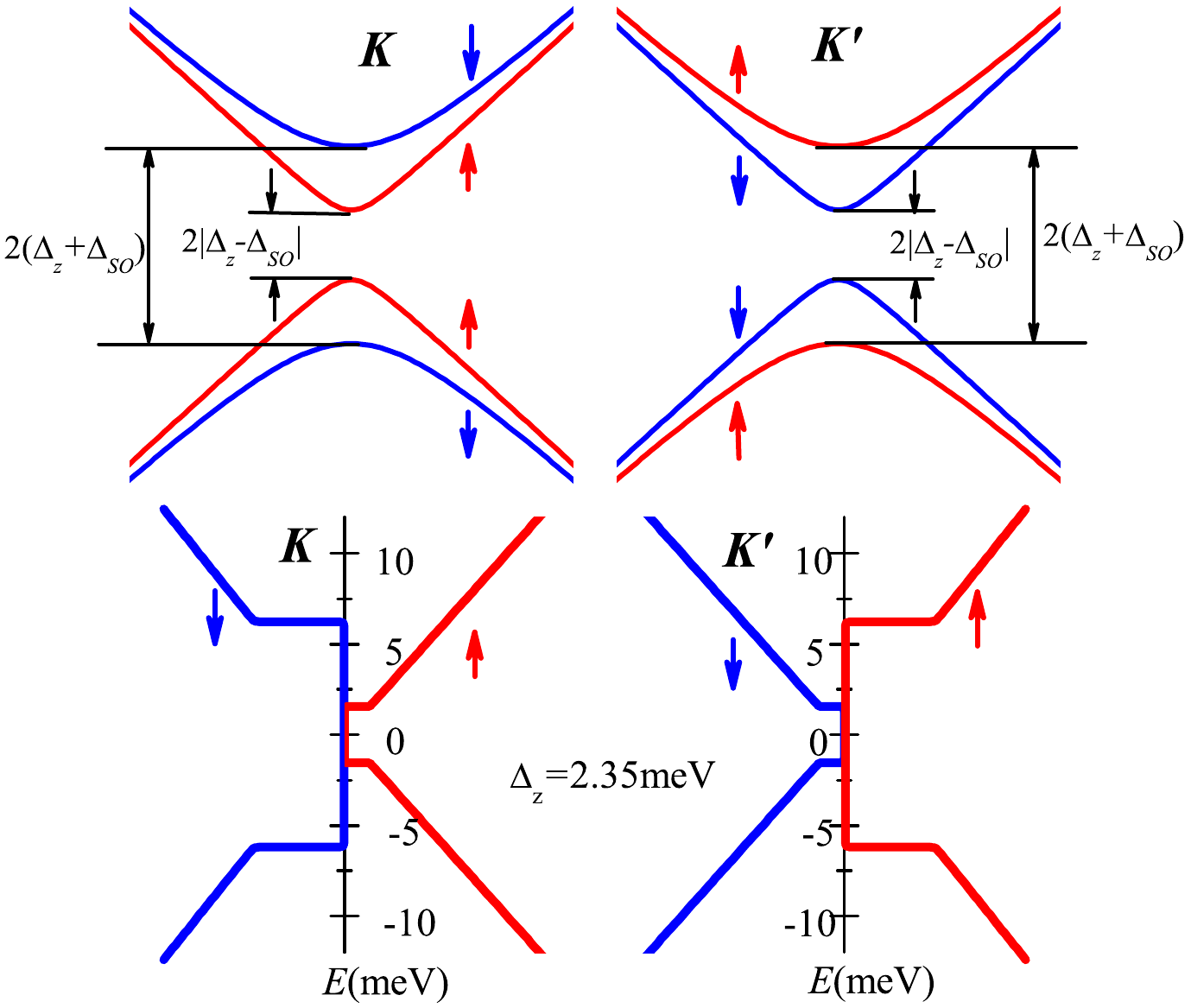}
   \caption{Sketches of (a) the dispersion relations and (b) densities of states of spin-up and spin-down electrons near Dirac nodes in pure silicene for $\Delta_z=2.35$\,meV. The arrows in (a) and (b) indicate the directions of electron spins.}  
      \label{DR-DOS}
  \end{figure}

The sketches of $\varepsilon_{\eta\sigma\mu}({\bf k})$ and of the DOSs of electrons in the pure system are given in Fig.\,\ref{DR-DOS}. It is clear that for Hamiltonian (1) there are four bands near each Dirac node, corresponding to the cases $\sigma=\pm 1$ and $\eta=\pm 1$. The values of energy gaps are $2|\Delta_z-\Delta_{SO}|$ and $2(\Delta_z+\Delta_{SO})$ for spin-up (spin-down) and spin-down (spin-up) bands near the $K$ ($K'$) node, respectively. The carriers near one Dirac node are spin polarized, but the system remains paramagnetic since spins of electrons near K and K' are polarized in opposite directions. From Fig.\,\ref{DR-DOS}(b), we also see that, in the pure silicene system, the DOSs of electrons linearly depend on $E$ when $E$ lies outside the energy gaps.

In realistic systems, the DOSs near the minima or maxima of bands strongly depend on
the electron-impurity scattering, which is usually described by a potential $V({\bf q})$ in the pseudo-spin basis. In pseudo-helicity basis, the scattering potential takes the form, $\hat V_{\eta\sigma;\mu\nu}({\bf k},{\bf k'})=\psi_{\eta\sigma\mu}^+({\bf k})V({\bf k}-{\bf k}')\psi_{\eta\sigma\nu}({\bf k}')$, corresponding to scattering of an electron in the valley $\eta$ with spin $\sigma$ from state $(\nu,{\bf k}')$ to state $(\mu,{\bf k})$ by impurities.

Further, we employ a Green's function approach to carry out the effects of electron-impurity interaction on density of states and optical conductivity in silicene. The previous studies of impurity problems in one-band models indicated that,\cite{PhysRevLett.48.886,PhysRevB.28.4704,PhysRevB.37.4589,PhysRevB.39.3400} to correctly describe the split-off impurity bands and the band tails, GF should be considered within the multiple-scatting approximation, first proposed by Klauder.\cite{Klauder196143} On the other hand, in silicene, the interband correlation induced by electron-impurity scattering is quite important: it leads to residual conductivity when the density of carriers in silicene essentially vanishes.\cite{PhysRevB.91.115411} Hence, to evaluate the Green's function,  generalizing the one-band multiple-scattering method to the two-band case with consideration of interband correlation is required.

In pseudo-helicity basis, the non-interacting retarded Green's function of an electron in valley $\eta$ with spin $\sigma$, $\hat g^r_{\eta\sigma;\mu\nu}({\bf k},E)$, takes a diagonal form ($\delta$ is an infinitesimal parameter)
\begin{align}
\hat g^r_{\eta\sigma;\mu\nu}({\bf k},E)=\frac{\delta_{\mu\nu}}{E-\varepsilon_{\eta\sigma;\mu}({\bf k})+i\delta},
\end{align}
while
the perturbative Green's function, $\hat G^r_{\eta\sigma;\mu\nu}({\bf k},E)$, relates to the the self-energy, $\hat \Sigma_{\eta\sigma;\mu\mu_1}^r({\bf k},E)$, via the Dyson's equation of the form
\begin{align}
  \hat G^r_{\eta\sigma;\mu\nu}({\bf k},E)=&\hat g_{\eta\sigma;\mu\mu}^r({\bf k},E)\delta_{\mu\nu}+\hat g_{\eta\sigma;\mu\mu}^r({\bf k},E)\nonumber\\
  &\times\hat \Sigma_{\eta\sigma;\mu\mu_1}^r({\bf k},E)\hat G^r_{\eta\sigma;\mu_1\nu}({\bf k},E).\label{eq3}
\end{align}
In the multiple-scattering approach, $\hat \Sigma_{\eta\sigma;\mu\mu_1}^r({\bf k},E)$ is determined by the Feynman diagrams presented in Fig.\,1. It can be written as
\begin{widetext}
\begin{align}
\hat \Sigma_{\eta\sigma;\mu\nu}^r({\bf k},E)=&n_i\sum_{{\bf q'},\mu_1\mu_2}\hat V_{\eta\sigma;\mu\mu_1}({\bf k},{\bf q}')\hat G_{\eta\sigma;\mu_1\mu_2}^r({\bf q}',E)\hat V_{\eta\sigma;\mu_2\nu}({\bf q}',{\bf k})\nonumber\\
&+n_i\sum_{\substack{{\bf q'},{\bf q}''\\ \mu_1\mu_2\mu_3\mu_4}}\hat V_{\eta\sigma;\mu\mu_1}({\bf k},{\bf q}')\hat G_{\eta\sigma;\mu_1\mu_2}^r({\bf q}',E)\hat V_{\eta\sigma;\mu_2\mu_3}({\bf q}',{\bf q}'')\hat G_{\eta\sigma;\mu_3\mu_4}^r({\bf q}'',E)\hat V_{\eta\sigma;\mu_4\nu}({\bf q}'',{\bf k})+...
\end{align}
with $n_i$ as the impurity density. Further, we introduce a vertex function, $\hat K_{\eta\sigma;\mu\nu}({\bf k},{\bf q};E)$, which satisfies the equation
\begin{align}
\hat K_{\eta\sigma;\mu\nu}({\bf k},{\bf q};E)=&\sum_{{\bf q'},\mu_1\mu_2}\hat V_{\eta\sigma;\mu\mu_1}({\bf q},{\bf q}')\hat G_{\eta\sigma;\mu_1\mu_2}^r({\bf q}',E)\left [n_i\hat V_{\eta\sigma;\mu_2\nu}({\bf q}',{\bf k})+\hat K_{\eta\sigma;\mu_2\nu}({\bf k},{\bf q}';E)\right ].\label{eq4}
\end{align}
\end{widetext}
Thus, we have $\hat \Sigma_{\eta\sigma;\mu\nu}^r({\bf k},E)=\hat K_{\eta\sigma;\mu\nu}({\bf k},{\bf q}={\bf k};E)$.

 \begin{figure}
   \centering
   \includegraphics[width=0.5\textwidth]{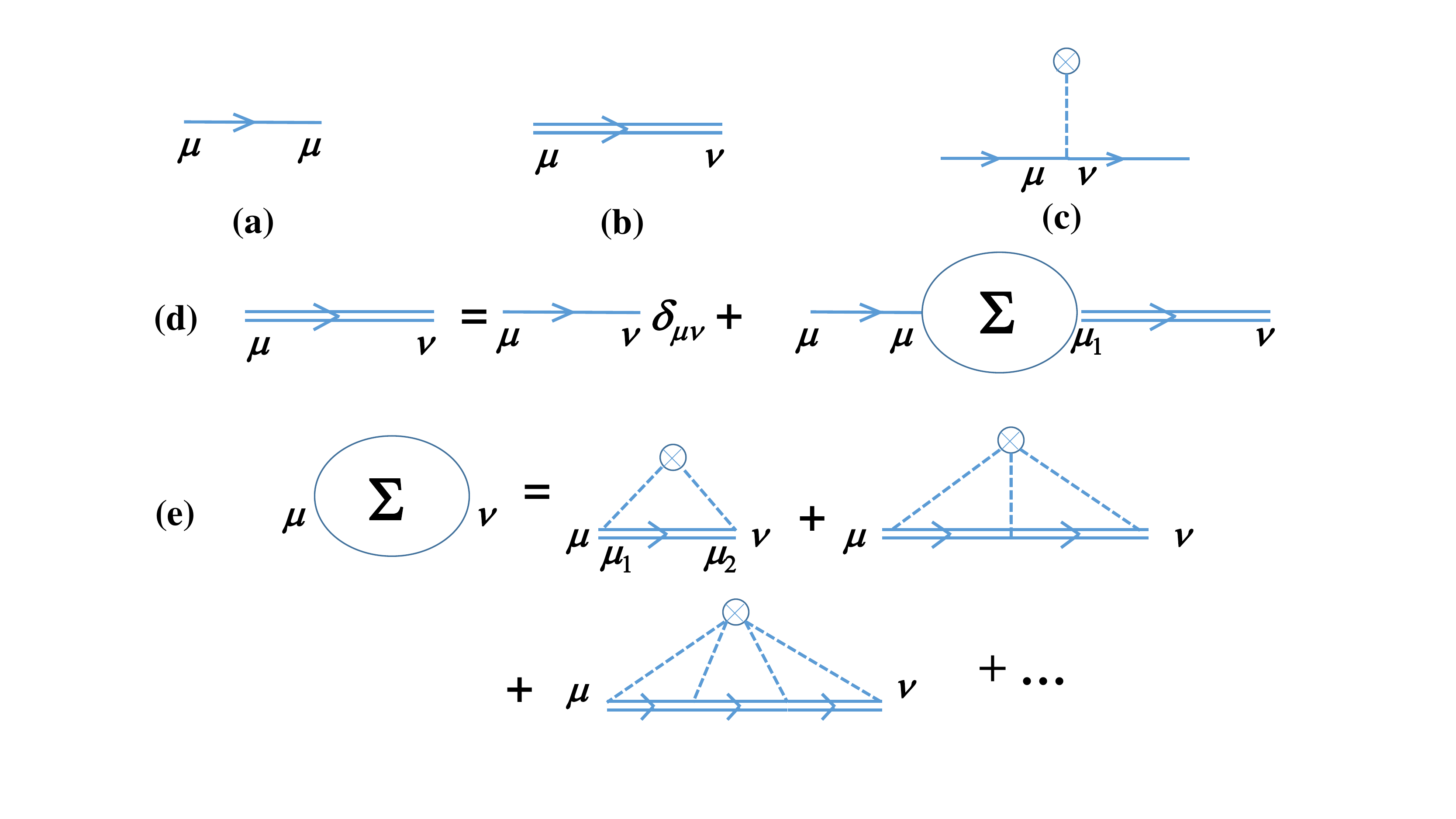}
   \caption{Feynman diagrams for evaluation of retarded Green's function within multiple-scattering approximation. (a), (b), and (c) show the non-interacting retarded GF, perturbative retarded GF, and the vertex of electron-impurity scattering, respectively. (d) is the Dyson's equation and (e) is the Feynman diagram of self-energy within multiple-scattering approximation. Here, the indices $\eta$ and $\sigma$ are dropped for brevity.}
   \label{Fig1}
 \end{figure}

 To solve Eqs.\,(\ref{eq3}) and (\ref{eq4}) in a self-consistent manner, we express perturbative retarded Green's function in terms of Fourier series, $\hat G_{\eta\sigma}^r({\bf k},\omega)=\sum_{n=0}^\infty \hat G_{\eta\sigma}^{c;nk}(\omega)\cos n\varphi_{\bf k}+\sum_{n=0}^\infty \hat G_{\eta\sigma}^{s;nk}(\omega)\sin n\varphi_{\bf k}$, and apply the iteration scheme proposed by Ng.\cite{jcp-61-2680} Whence $\hat G_{\eta\sigma}^r({\bf k},\omega)$ is carried out, the density of states of electrons with spin $\sigma$ near node $\eta$, defined as $D_{\eta\sigma}(\omega)=\sum_{{\bf k},\mu}\hat G^r_{\eta\sigma;\mu\mu}({\bf k},\omega)$, can be obtained directly from the zeroth term of cosine Fourier series: $D_{\eta\sigma}(\omega)=\sum_{{\bf k},\mu}\hat G^{c;0k}_{\eta\sigma;\mu\mu}(\omega)$.

 In experiment, the frequency-dependent optical conductivity is a powerful probe to measure the electronic states in materials. Ignoring the influence of electronic states induced by electron-impurity scattering, the ac conductivity in silicene has been investigated by Vargiamidis {\it et al}.\cite{0953-8984-26-34-345303} In Ref.\,[\onlinecite{1367-2630-16-10-105007}], the optical properties beyond the usual Dirac-cone approximation in {clean} silicene were also studied by first-principle calculation.
 Considering the change of band structure due to defects, the optical conductivity in silicene has been carried out recently.\cite{Zakerian2016} In these studies, the Kubo formula based on single-particle assumption was employed and the interband coherence was completely ignored. In present paper, we generalize the Kubo formula in the presence of the interband correlation to investigate the optical conductivity in silicene (the detailed procedure of deriving the Kubo formula is presented in Appendix). Ignoring the vertex correction, real part of zero-temperature longitudinal conductivity for electrons with spin $\sigma$ near Dirac node $\eta$, ${\rm Re}\sigma_{\eta\sigma;xx}(\omega_0)$, takes the form
\begin{widetext}
\begin{align}
  {\rm Re}\sigma_{\eta\sigma,xx}(\omega_0) =&\frac{1}{\omega_0}\sum_{\substack{\mu,\nu \\ \mu_1,\nu_1,{\bf k}}}\int_{\mu-\omega_0}^\mu \frac{d\omega_1}{2\pi}\left\{ {\rm Re}[\hat j_{\eta\sigma;\nu_1\mu}^x({\bf k})\hat j_{\eta\sigma;\mu_1\nu}^x({\bf k})+\hat j_{\eta\sigma;\nu\mu}^x({\bf k})\hat j_{\eta\sigma;\mu_1\nu_1}^x({\bf k})][{\rm Im}\hat {G}^{c;0k}_{\eta\sigma;\mu\mu_1}{\rm Im}{ G}^{c;0k}_{\eta\sigma;\nu_1\nu}]_{\omega_1+\omega_0,\omega_1}\right .\nonumber\\
&\left . {\rm Re}[\hat j_{\eta\sigma;\nu_1\mu}^x({\bf k})\hat j_{\eta\sigma;\mu_1\nu}^x({\bf k})-\hat j_{\eta\sigma;\nu\mu}^x({\bf k})\hat j_{\eta\sigma;\mu_1\nu_1}^x({\bf k})][{\rm Re}\hat {G}^{c;0k}_{\eta\sigma;\mu\mu_1}{\rm Re}{ G}^{c;0k}_{\eta\sigma;\nu_1\nu}]_{\omega_1+\omega_0,\omega_1}
  \right \}.\label{eq5}
\end{align}
\end{widetext}
Here, $\hat j^{i}_{\eta\sigma;\mu\nu} ({\bf k})$ are the elements of $i-$th ($i=x,\,\,y$) component of  single-particle current  in pseudo-helicity basis and take the forms $\hat j^{i}_{\eta\sigma;\mu\nu} ({\bf k})=-ev_F\psi^+_{\eta\sigma\mu}({\bf k})\left(\frac{\partial \check h_{\eta\sigma}({\bf k})}{\partial k_i}\right)_{\mu\nu}\psi_{\eta\sigma\nu}({\bf k})$. To derive Eq.\,(\ref{eq5}), we assume that the dominant contribution to current comes from the zeroth-order term of Fourier series of $\hat G^r$ and the contribution associated with higher-order terms is ignored. 

\section{Numerical Results}

Further, we present a numerical calculation to investigate the effects of electron-impurity interaction on the density of states and on the optical conductivity in silicene. In calculation, the characteristic energy of SO coupling due to external electric field is chosen as $\Delta_{z}=2.35$\,meV. We assume that the main contribution to self-energy of electrons comes from a screened scattering potential due to charged impurities: $V(q)=e^2/[2\kappa\epsilon_0q\epsilon(q)]$. Here, $\kappa$ is the dielectric constant of substrate and  $\epsilon(q)=1-v(q)\Pi(q)$ is the static dielectric function.  $\Pi(q)$ is the static polarization function. At zero temperature, it takes the form\cite{PhysRevD.33.3704,PhysRevB.78.075433,0953-8984-21-7-075303,0953-8984-21-2-025506,1402-4896-83-3-035002,PhysRevB.89.195410,PhysRevB.90.035142}
\begin{align*}
\Pi(q)&=-\frac{\mu_0}{2\pi v_F^2}\sum_{\sigma,\eta}\left [F(q)\theta(|\lambda_{\eta\sigma}|-\mu_0)\right .\nonumber\\
&\left . +G(q)\theta(\mu_0-|\lambda_{\eta\sigma}|)\right],
\end{align*} 
with $\mu_0$ as the chemical potential. $F(q)$ and $G(q)$, respectively, take the forms [$k_F^{\eta\sigma}=\sqrt{\mu_0^2-\lambda_{\eta\sigma}^2}$]
\begin{equation}
F(q)=\frac{|\lambda_{\eta\sigma}|}{2\mu_0}+\frac{v_F^2q^2-4\lambda_{\eta\sigma}^2}{4v_Fq\mu_0}\arcsin\left(\sqrt{\frac{v_F^2q^2}{v_F^2q^2+4\lambda_{\eta\sigma}^2}}\right)
\end{equation}
and
\begin{align*}
&G(q)=1-\theta(q-2k_F^{\eta\sigma})\left[\frac{\sqrt{q^2-4(k_F^{\eta\sigma})^2}}{2q}\right .\nonumber\\
&\left .-\frac{v_F^2q^2-4\lambda_{\eta\sigma}^2}{4v_Fq\mu_0}\arctan\left({\frac{\sqrt{v_F^2q^2-4(k_F^{\eta\sigma})^2v_F^2}}{2\mu_0}}\right)\right ].
\end{align*}

\subsection{Split-off impurity bands and band-gap narrowing}

We first analyze the effects of electron-impurity scattering on density of states in silicene. It is well known that in conventional semiconductors, introduction of impurities produces local energy levels lying within the energy gap. These levels broaden into impurity bands when the concentration of impurities increases. Further, in heavily doped semiconductors, the impurity bands may be combined with a conduction or valence band, forming a band tail and leading to band-gap narrowing. In conventional two-dimensional electron gases, the transition from the split-off impurity band at low impurity concentration to a band tail at high impurity concentration has also been demonstrated theoretically.\cite{PhysRevB.37.4589,PSSB:PSSB375,apl-102-162101} Hence, the similar phenomena are expected to be observed in silicene.

\begin{figure}
   \centering
   \includegraphics[width=0.5\textwidth]{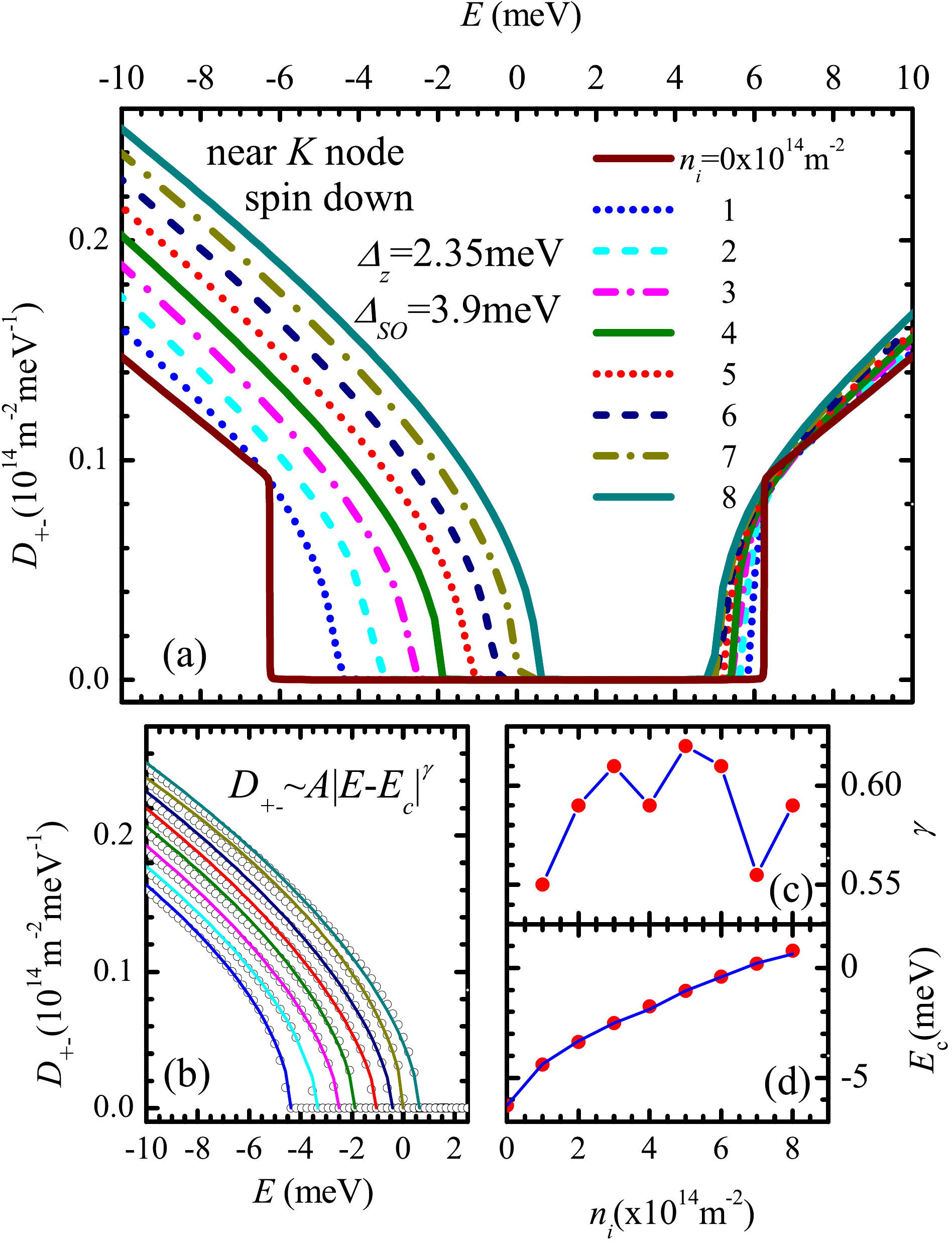}
   \caption{(a) Densities of states of spin-down electrons in silicene for various impurity densities near Dirac point $K$. The chemical potential is $\mu_0=20$\,meV and the energy due to the SO coupling induced by the external electric field is $\Delta_z=2.35$\,meV. (b) Fitting of DOSs near the top of the lower band in (a) by the power-law formula $D_{\pm}=A|E-E_c|^\gamma$.  The data from (a) are shown by open circles and the solid lines are the best fit lines of these data. From bottom to top, the impurity densities are $n_i=1,\,\,2,\,\,3,\,\,4,\,\,5,\,\,6,\,\,7$, and $8\times 10^{14}$\,m$^{-2}$. The dependencies of the fitting parameters $\gamma$ and $E_c$ on impurity densities are shown by filled circles in (c) and (d), respectively. The solid line in (d) is the best fit line of data: $\frac{E_c}{{\rm meV}}=-5.95+1.4\left (\frac{n_i}{10^{14}{\rm m}^{-2}}\right )^{0.75}$.}   
         \label{FigNBPM}
\end{figure}

In Fig\,\ref{FigNBPM}(a) we plot the densities of states of electrons with spin down near the $K$ node ({\it i.e.} $\sigma=-1$ and $\eta=1$) for various impurity densities. The chemical potential is $\mu_0=20$\,meV. It is clear that, when the impurity density increases, the energy gaps become narrower. In our study, a repulsive potential of electron-impurity scattering is considered and the impurities essentially play roles as acceptors. Hence, band-gap narrowing mainly comes from the movement of top of the lower energy band towards the high-energy side. However, remarkable shifts of upper energy band bottom towards the low-energy side also can be observed due to strong band correlation in silicene.

\begin{figure}
   \centering
   \includegraphics[width=0.5\textwidth]{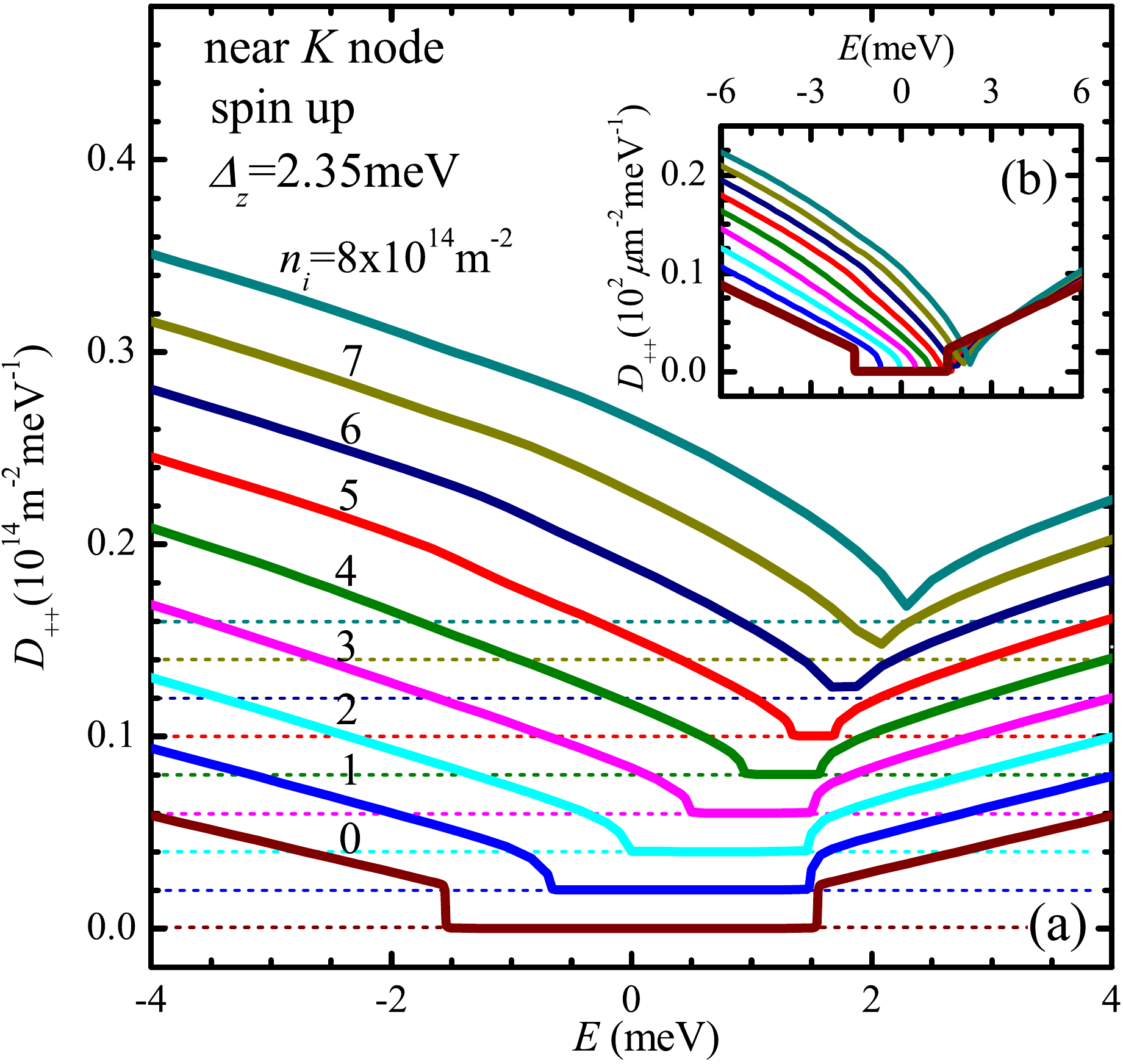}
   \caption{(a) Densities of states of spin-up electrons in silicene for various impurity densities near the $K$ node. For a guide of eyes, each curve moves towards the upper side by $\Delta D_{++}=2\times 10^{12}$\,m$^{-2}$meV$^{-1}$ in sequence and the positions of $D_{++}=0$ are indicated by horizontal dashed lines. (b) The same $D_{xx}$ versus $E$ as in (a) but without $\Delta D_{++}$ shift.}  
      \label{FigNB}
\end{figure}

Note that when $n_i=0$, there are two discontinuities in $D_{\pm}$ versus $E$ at $E=\pm |\Delta_{SO}-\Delta_z|$. In the presence of electron-impurity scattering, they are smeared out and $D_\pm$ continuously changes with $E$. Such $D_\pm$ versus $E$ can be described by a power-law formula, $D_{\pm}\sim A|E-E_c|^\gamma$. In Fig.\,\ref{FigNBPM}(b) we show the fitting of $D_\pm$ near the top of the lower energy band for various impurity densities. The parameters $\gamma$ and $E_c$ are assumed to be $n_i$-dependent: the $\gamma$ and $E_c$ versus $n_i$ are given by Figs.\,\ref{FigNBPM}(c) and \ref{FigNBPM}(d), respectively. We find that the values of $\gamma$ are between $0.5\sim 0.65$ and the $E_c$ versus $n_i$ can be further fitted by $\frac{E_c}{{\rm meV}}=-5.95+1.4\left (\frac{n_i}{10^{14}{\rm m}^{-2}}\right )^{0.75}$.   

In Fig.\,\ref{FigNBPM}, the spins of carriers near $K$ node are assumed to be polarized in the down direction. Correspondingly, the energy gap is 2$(\Delta_z+\Delta_{SO})=10.5$\,meV, which is relatively large. Hence, in the case $n_i\le 8\times 10^{14}$\,m$^{-2}$, the complete disappearance of energy gaps does not be observed. However, for the spin-up electrons near $K$ node, the energy gap becomes small: 2$|\Delta_z-\Delta_{SO}|=3.1$\,meV and the vanishing of energy gap is relatively easy to be seen. In Fig.\,\ref{FigNB}, we plot the energy dependencies of densities of states of electrons with spin-up near $K$ node ({\it i.e.} $\eta=+$ and $\sigma=+$). It is clear that when the density of impurities increases from $n_i=0$ but still remains relatively low, band-gap narrowing can be observed. But when $n_i$ further ascends that $n_i \gtrsim 5\times 10^{14}$\,m$^{-2}$, the energy gap disappears completely. 

\begin{figure}
  \centering
  \includegraphics[width=0.5\textwidth]{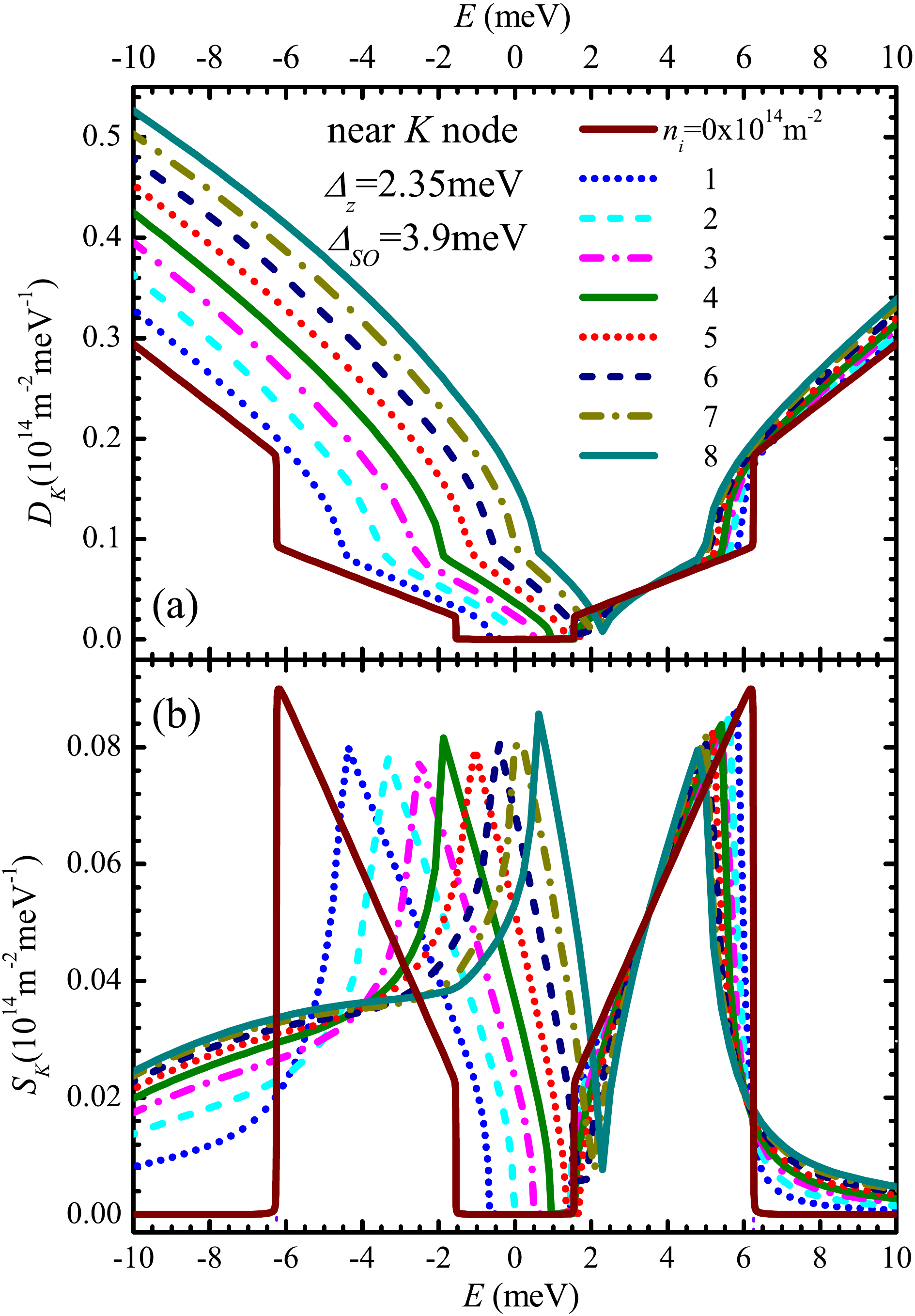}
  \caption{Energy dependencies of total DOSs, $D_{K}=D_{++}+D_{+-}$ (a), and of spin-polarized DOSs, $S_K=D_{++}-D_{+-}$ (b), of electrons near the $K$ node for various impurity densities. The other parameters are the same as in Fig.\,\ref{FigNBPM}}.
  \label{FigNBtotal}
\end{figure}

Further, in Fig.\,\ref{FigNBtotal}(a), we plot the energy-dependencies of total DOSs of electrons near $K$ node, $D_K=D_{++}+D_{+-}$, for various impurity densities. In the absence of electron-impurity scattering, four discontinuities in $D_K$ versus $E$ can be observed at $E=\pm |\Delta_{SO}\pm \Delta_z|$. In the presence of impurities, they are smeared out but "dog-leg" shaped connections still can be observed. When energy increases, $D_K$ first decreases and then it may reach the zero value for $n_i\lesssim 5\times 10^{14}$\,m$^{-2}$, forming an energy gap. When $E$ further increases, $D_K$ also increases. When impurity density is relatively high ($n_i\gtrsim 5\times 10^{14}$\,m$^{-2}$), $D_K$ is always nonvanishing and the energy gap disappears completely.  

One of interesting properties in silicene is that the electrons near each Dirac node are spin-polarized, although the net spin-polarization of electrons vanishes. In Fig.\,\ref{FigNBPM}(b), we show the energy dependencies of spin-polarized DOSs, defined as $S_{\eta}=D_{\eta +}-D_{\eta -}$,  for various densities of impurities. We see that in pure silicene, $S_K$ is nonvanishing only in the energy ranges $-|\Delta_{SO}+\Delta_z|<E<-|\Delta_{SO}-\Delta_z|$ (denoted as range I) and $|\Delta_{SO}-\Delta_z|<E<|\Delta_{SO}+\Delta_z|$ (denoted as range II). When $E$ increases, $S_K$ linearly decreases in range I while it linearly increases in range II. In the presence of electron-impurity scattering, the range of nonvanishing $S_K$ becomes broader. In particular, when $n_i$ increases, the range of vanishing of $S_K$ between $-|\Delta_{SO}-\Delta_z|<E<|\Delta_{SO}-\Delta_z|$ becomes narrower and finally disappears if $n_i \gtrsim 5\times 10^{14}$\,m$^{-2}$. Besides, $S_K$ is also nonvanishing for $|E|>|\Delta_{SO}+\Delta_z|$ in the presence of impurities. 

\begin{figure}
   \centering
   \includegraphics[width=0.5\textwidth]{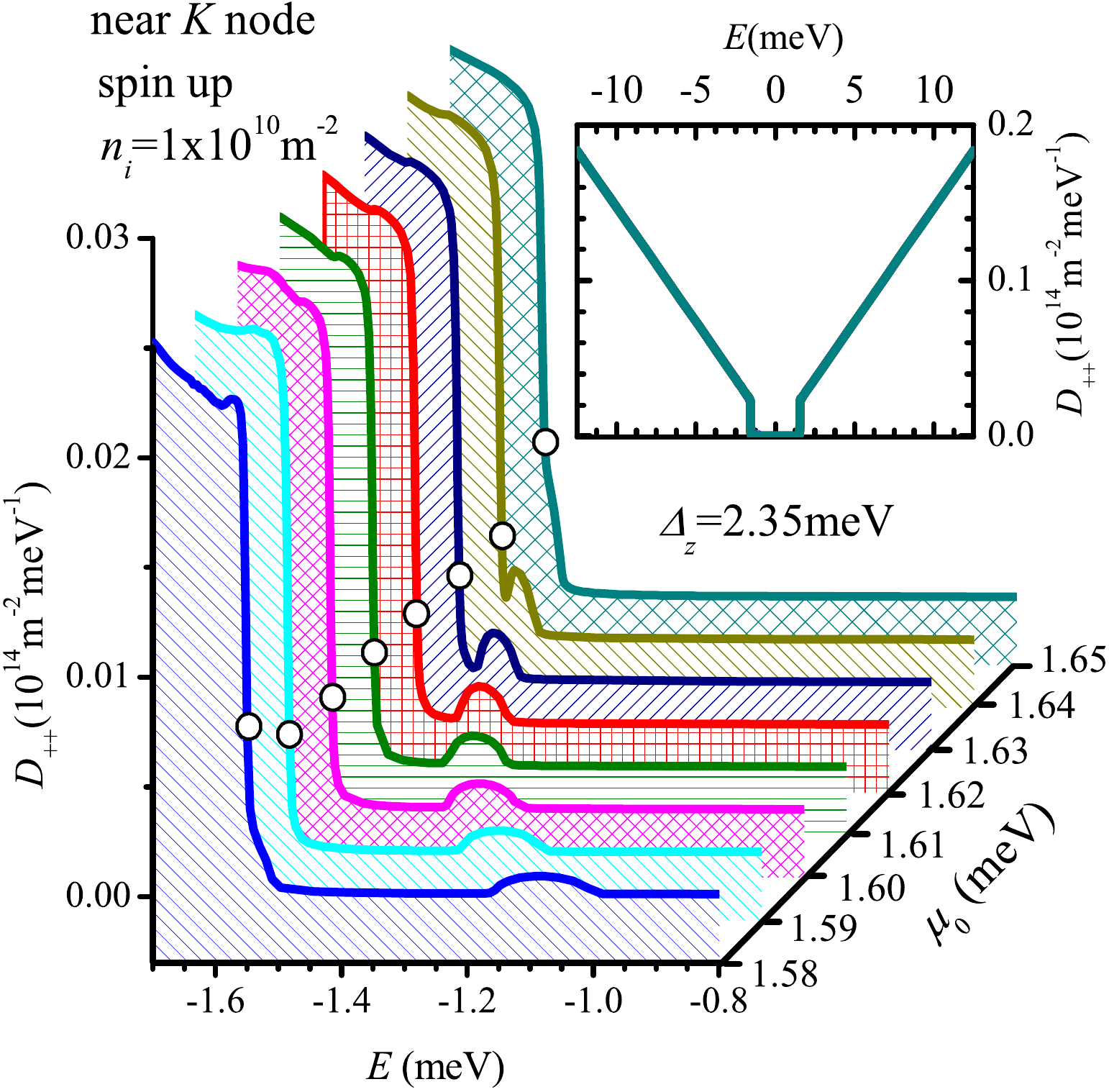}
   \caption{DOSs of spin-up electrons near $K$ node in silicene for various chemical potentials near the top of the upper energy band. The density of impurities is $n_i=1\times 10^{-10}$\,m$^{-2}$ and $\Delta_z=2.35$\,meV. From bottom to top, the chemical potentials are $\mu_0=1.58$, $1.59$, $1.60$, $1.61$, $1.62$, $1.63$, $1.64$, and $1.65$\,meV, correspondingly.  The circles on curves indicate the DOSs at energy $E=-1.55$\,meV, which corresponds to the lower limit of energy gap for electrons with $\eta=\sigma=+1$ in the pure system. The inset shows $D_{++}$ versus $E$ for various chemical potentials within an  enlarged energy scale.}  
      \label{FigIB}
  \end{figure}

In Figs.\,\ref{FigNBPM}-\ref{FigNBtotal} we do not observe the split-off impurity bands since in these cases the chemical potential is relatively large. To demonstrate the IBs, in Fig.\,\ref{FigIB}, we plot the densities of states of electrons with spin-up near $K$ node for various chemical potentials closed to upper limit of gap $|\Delta_{SO}-\Delta_z|=1.55$\,meV: $\mu_0=1.58$, $1.59$, $1.60$, $1.61$, $1.62$, $1.63$, $1.64$, and $1.65$\,meV. The concentration of impurity is relatively dilute, $n_i=1\times 10^{10}$\,m$^{-2}$. It is obvious that, for these chemical potentials, $D_{++}$ are almost the same within the most part of energy range studied here (see the inset of Fig.\,\ref{FigIB}), except for $E$ near the top of lower energy band. We see that in the case $E$ closed to $-1.55$\,meV, {\it .i.e.} the lower limit of energy gap of pure silicene, impurity bands are formed for $\mu_0=1.58\sim 1.64$\,meV. When the chemical potential increases, the center of IB moves towards the low-energy side that it finally disappears and combines with the valence bands for $\mu_0=1.65$\,meV.

\subsection{Optical conductivity}

After self-consistent evaluation of retarded Green's functions within the multiple-scattering approximation, the optical conductivity can be carried out by means of Eq.\,(\ref{eq5}). The results are presented in Figs.\,7 and 8.

\begin{figure}
   \centering
   \includegraphics[width=0.5\textwidth]{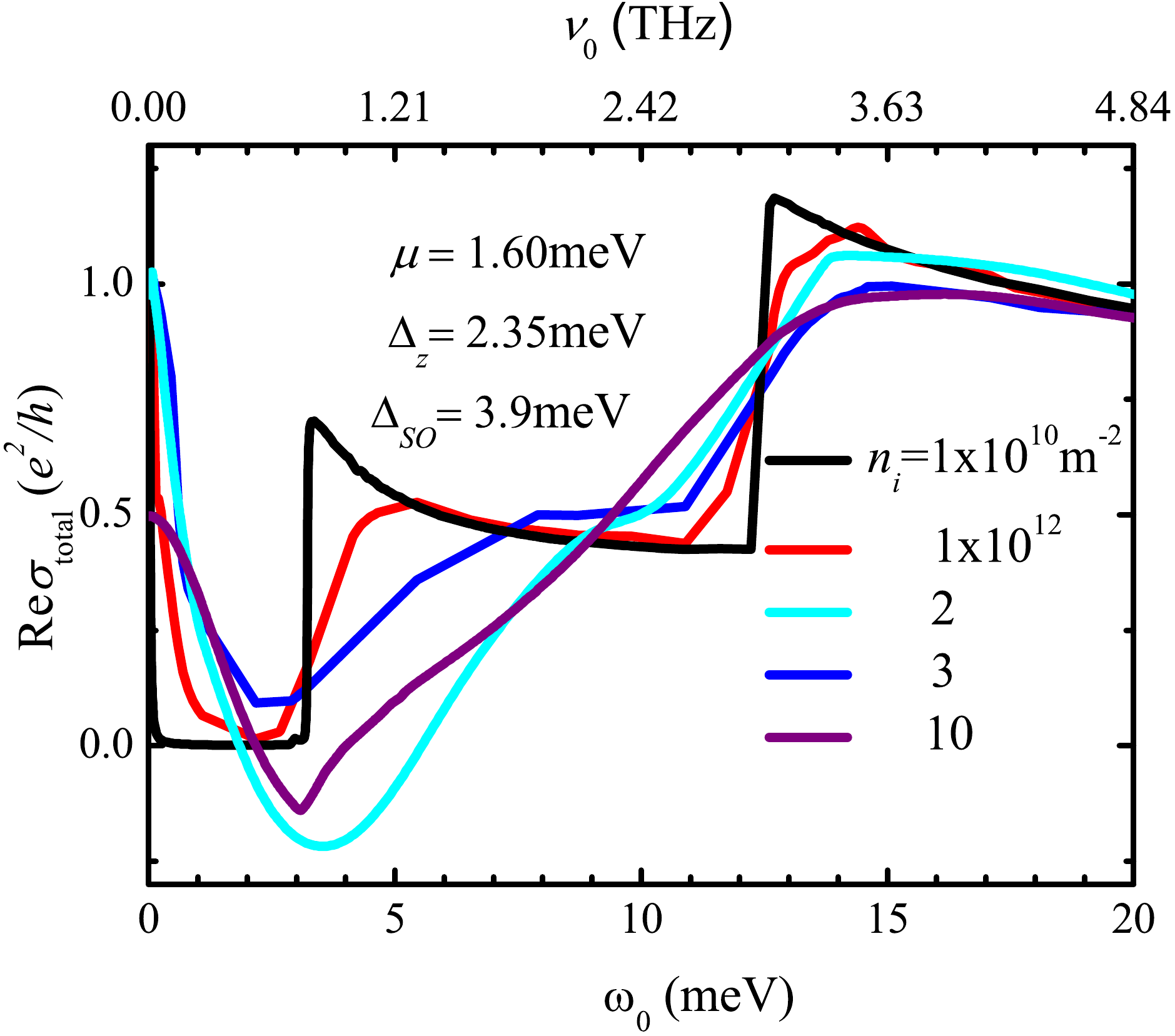}
   \caption{The frequency dependencies of real part of total optical conductivity in silicene for various impurity densities $n_i=1\times 10^{10}$, $1\times 10^{12}$, $2\times 10^{12}$, $3\times 10^{12}$, and $1\times 10^{13}$\,${\rm m}^{-2}$. The chemical potential is $\mu_0=1.60$\,meV, $\Delta_z=2.35$\,meV, and $\Delta_{SO}=3.9$\,meV.}
      \label{FigOC1}
\end{figure}
In Fig.\,7, we plot the real part of total optical conductivity, ${\rm Re}\sigma_{\rm total}\equiv \sum_{\eta,\sigma=\pm}{\rm Re}\sigma_{\eta\sigma,xx}(\omega_0)$, versus frequency $\omega_0$ (or $\nu_0=\omega_0/h$) for various impurity densities $n_i=1\times 10^{10}$, $1\times 10^{12}$, $2\times 10^{12}$, $3\times 10^{12}$, and $1\times 10^{13}$\,${\rm m}^{-2}$. The chemical potential is chosen to be $\mu_0=1.60$\,meV, which is closed to the upper limit of energy gap. When the impurity density is relatively small (in the case $n_i=1\times 10^{-10}$\,${\rm m}^{-2}$), we can observe two peaks which correspond to the optical excitations of electrons from two branches of valence bands: when $\omega_0$ increases from $\omega_0<2|\Delta_z\pm \Delta_{SO}|$ ({\it i.e.} $3.1$ and $6.5$\,meV) to  $\omega_0>2|\Delta_z\pm \Delta_{SO}|$, ${\rm Re}\sigma_{\rm total}$ first abruptly increases and then gradually decreases. When the impurity density increases and reaches the value of order of $1\times 10^{12}$\,${\rm m}^{-2}$, two peaks begin to be smeared out due to the increase of DOS within the energy gap.  The peak near $\omega_0=6.5$\,meV reduces monotonically when $n_i$ ascends. However, near the lower frequency ({\it i.e.} near $\omega_0=2|\Delta_z- \Delta_{SO}|=3.1$\,meV) the peak first decreases rapidly when $n_i$ increases and it disappears completely in the case $n_i=2\times 10^{12}$\,${\rm m}^{-2}$. When $n_i$ further ascends, the dependencies of ${\rm Re}\sigma_{\rm total}$ on $n_i$ for $\omega_0$ near $2|\Delta_z- \Delta_{SO}|$ become non-monotonic: ${\rm Re}\sigma_{\rm total}$ first increases when $n_i$ increases from $2$ to $5\times 10^{12}$\,${\rm m}^{-2}$ and it reduces with a further increase of $n_i$. 

In Fig.\,8 we plot the frequency dependencies of real part of total optical conductivity for the chemical potential $\mu_0=20$\,meV, which corresponds to the metallic phase of silicene. When the frequency increases, ${\rm Re}\sigma_{\rm total}$ decreases monotonically. The dependencies of real part of optical conductivity on impurity density are quite distinct for high and low optical frequencies: when $n_i$ increases, ${\rm Re}\sigma_{\rm total}$ decreases for $\omega_0<0.5$\,meV, while it increases in the case $\omega_0>2.5$\,meV.

\begin{figure}
   \centering
   \includegraphics[width=0.5\textwidth]{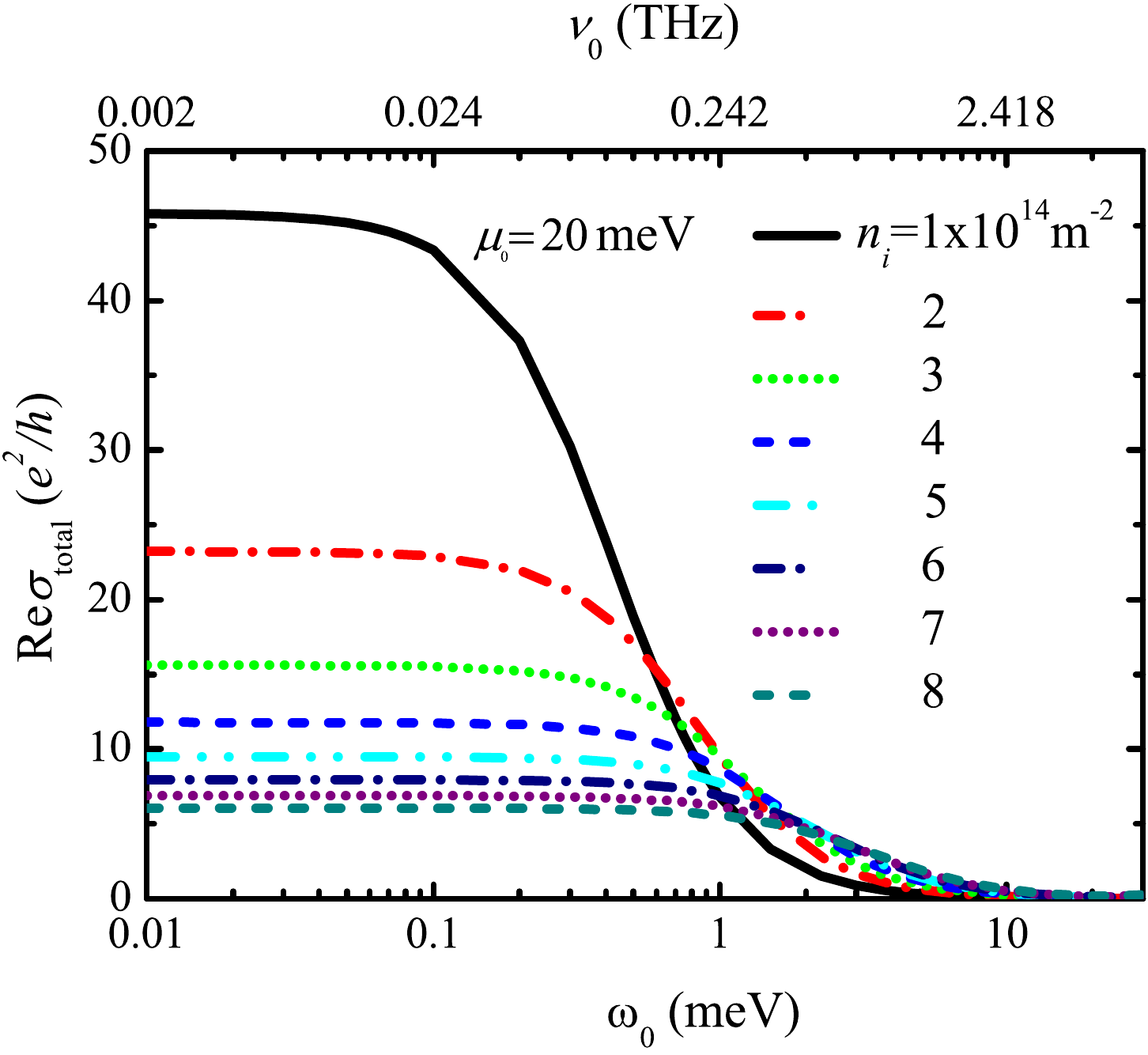}
   \caption{${\rm Re}\sigma_{\rm total}$ versus $\omega_0$ (or $\nu_0$) for various impurity densities $n_i=1$, $2$, $3$, $4$, $5$, $6$, $7$, $8\times 10^{14}$\,${\rm m}^{-2}$. The chemical potential is $\mu_0=20$\,meV. Other parameters are the same as those in Fig.\,\ref{FigOC1}.}
      \label{FigOC2}
  \end{figure}

From Figs.\,7 and 8, it is clear that, to observe band-gap narrowing in the optical conductivity versus frequency, the chemical potential of silicene samples should be closed to the energy gap and the concentration of impurities should be relatively dilute: $n_i$ is of order of $10^{12}$\,m$^{-2}$. Besides, we also clarify that it is difficult to detect the split-off impurity bands from the optical conductivity study since the DOSs of impurity bands are much smaller than those out of energy gaps. To observe the IBs, more powerful experimental tools, such as angle-resolved photoemission spectroscopy (ARPES), pump-probe spectroscopy within terahertz regime, {\it etc}. are required.   

\section{Conclusions}

Generalizing the multiple-scattering approach of Green's function to consider the interband correlation, the effects of electron-impurity scattering on the density of states have been investigated. We find that, in the case of relatively high chemical potential, the energy gap reduces with an increase of impurity density and it finally disappears when $n_i$ reaches the magnitude of order of $\sim 5\times 10^{14}$\,${\rm m}^{-2}$. The split-off impurity bands can be observed only for low CP and low $n_i$. These bands transform into the band tails as the CP (or $n_i$) increases. We also find that, in the frequency dependencies of real part of conductivity for low $n_i$ and low CP, there are two peaks which correspond the interband excitations of electrons. These peaks are smeared out when the impurity density ascends. In the case of high chemical potential, ${\rm Re}\sigma_{xx}$ versus $n_i$ shows distinct behaviors for low and high frequency: as the impurity density increases, ${\rm Re}\sigma_{xx}$ decreases for low $\omega_0$ while it increases in the case of high $\omega_0$. 

\begin{acknowledgments}
This work was supported by the project of National Key Basic Research
Program of China (973 Program) (Grant No. 2012CB927403) and National
Natural Science Foundation of China (Grant No. 11274227). 
\end{acknowledgments}

\appendix*

\section{generalized Kubo formula in the presence of interband correlation}

In the previous studies on linear multiband transport, Kubo formula without the vertex corrections has been widely used to interpret the dc and ac conductivities. In the typical form of this formula, the effect of interband transition induced by external dc and/or ac electric fields is considered, but the interband correlation induced by electron-impurity scattering usually is ignored. However, the last one is quite important in the narrow-band semiconductors as well as in new-type two-dimensional systems, such as graphene, silicene, and germanene {\it etc}. Hence, to correctly describe the dc and ac transport properties in these systems, a new generalized Kubo formula is required.

We consider an equilibrium system of carriers, which may be scattered by impurities, phonons, {\it etc}. The single-particle Hamiltonian in the absence of scatterings is denoted by $\hat h_0(\hat{\bf p})$ with $\hat {\bf p}$ as the carrier momentum operator. Further, we assume that the eigenfunctions of $\hat h_0(\hat {\bf p})$ are known: they are denoted by $\psi_i({\bf r})$ with $i$ as the index of eigenvalues $E_i$. In the framework of Green's function approach, the motion of such an equilibrium system can be determined by the GFs, $\hat G^{r,a,<}({\bf r},t;{\bf r}',t')$, in which the scatterings of carriers due to impurities, phonons, {\it etc.} are embedded. In the basis of eigenfunctions $\psi_i$, they can be rewritten as 
\[
\hat G^{r,a,<}({\bf r},t;{\bf r}',t')=\sum_{i,j}\hat G^{r,a,<}_{ij}(t,t')\psi_i({\bf r})\psi^*_j({\bf r}'),
\]
where $\hat G^{r,a,<}_{ij}(t,t')$ are the GFs based on eigenfunctions of $\hat h_0$ and they essentially depend only on the difference of two times. Note that according to Kubo-Martin-Schwinger relations,\cite{PhysRev.115.1342,doi:10.1143/JPSJ.12.570}  $G^<$ relates to $G^{r,a}$ in the $\omega$ space by
\[
\hat{G}^<_{ij}(\omega)=-n_{\rm F}(\omega)[\hat G^r_{ij}(\omega)-\hat G^a_{ij}(\omega)].
\]

Further, we assume that the system is driven by an external electric field ${\bf E}(t)$, described  by the vector potential ${\bf A}(t)$. In the framework of minimum coupling, the single-particle non-interacting Hamiltonian takes the form $\hat h'_0\equiv\hat h_0(\hat {\bf p}+e{\bf A}(t))$. Up to the first order of electric field,  $\hat h_0'$
can be further rewritten as $\hat h_0'\approx\hat h_0(\hat{\bf p})+\delta \hat h_0$. Here, $\delta h_0$ is  the perturbed part due to ${\bf E}(t)$ and takes the form $\delta \hat h_0=-\hat{\bf j}_0({\bf r}) \cdot {\bf A}(t)$ with $\hat {\bf j}_0({\bf r})=-e{\bm \nabla}_{\bf p}\hat h_0(\hat{\bf p})$. Thus, up to the first order of ${\bf A}(t)$, the lesser Green's function out of equilibrium, $\hat {\cal G}^<$, takes the form
\begin{widetext}
\begin{align}
\hat {\cal G}^<({\bf r},t;{\bf r}',t')=\hat {G}^<({\bf r},t;{\bf r}',t')-\frac 12\int d{\bf r}''\int dt\hat {G}({\bf r},t;{\bf r}'',t'')\{{\bf A}(t'')\cdot[\overrightharpoon{\hat {\bf j}_0}({\bf r}'')+\overleftharpoon{\hat {\bf j}^+_0}({\bf r}'')]\}\hat { G}({\bf r}'',t'';{\bf r}',t')]^<,\label{A1}
\end{align}
\end{widetext}
where the symbol $\rightharpoonup$ or $\leftharpoonup$, standing over the operators, denotes the direction of action.  

To evaluate the conductivity, one has to carry out the single-particle current operator, ${\bf {j}}$. From the definition of current, ${\bf { j}}=-e\frac{d{\bf r}}{dt}$, and the motion of equation of ${\bf r}$, it follows
\begin{align}
  \hat {\bf {j}}({\bf r},t)=-ie[\hat h_0',{\bf r}]
  \approx -e {\bm \nabla}_{\bf p} \hat h_0({\bf p})-e^2[{\bf A}(t)\cdot {\bm \nabla}_{\bf p}]{\bm \nabla}_{\bf p}\hat h_0({\bf p}).\label{A2}
\end{align}
$\hat {\bf {j}}({\bf r},t)$ can be further rewritten as $\hat {\bf {j}}({\bf r},t)=\hat {\bf j}_0({\bf r})+\delta \hat {\bf { j}}({\bf r},t)$, with $\delta \hat {\bf { j}}({\bf r},t)\equiv -e^2[{\bf A}(t)\cdot {\bm \nabla}_{\bf p}]{\bm \nabla}_{\bf p}\hat h_0(\hat{\bf p})$. The macroscopic current that observed in experiments, ${\bf J}({\bf r},t)$, is determined by $J({\bf r})=-i\lim_{{\bf r'}\rightarrow {\bf r}}\frac 12 [\hat {\bf { j}}({\bf r},t)+\hat {\bf { j}^+}({\bf r}',t)]{\cal G}^<({\bf r},t;{\bf r}',t)$. Substituting Eqs.\,(\ref{A1}) and (\ref{A2}) into ${\bf J}({\bf r},t)$ and retaining the terms up to the first order of ${\bf E}(t)$, we get
\begin{widetext}
  \begin{align}
    {\bf J}({\bf r},t)=&-\frac i2\lim_{{\bf r}'\rightarrow{\bf r}}[\delta \hat {\bf j}({\bf r},t)+\delta \hat {\bf j}^+({\bf r}',t)]\hat {G}^<({\bf r},t;{\bf r}',t)\nonumber\\
    &+\frac i{4}\lim_{{\bf r}'\rightarrow {\bf r}}[\hat {\bf j}_0({\bf r})+\hat {\bf j}_0^+({\bf r}')]\int d{\bf r}''\int dt''[\hat {G}({\bf r},t;{\bf r}'',t'')\{{\bf A}(t'')\cdot[\overrightharpoon{\hat {\bf j}_0}({\bf r}'')+\overleftharpoon{\hat {\bf j}_0^+}({\bf r}'')]\}\hat { G}({\bf r}'',t'';{\bf r}',t)]^<.
    \end{align}
    In the terms of eigenfunction representation of ${G}^<$, ${\bf J}({\bf r},t)$ can be further rewritten as
  \[
    {\bf J}({\bf r},t)=-i\sum_{ij}\delta {\bf j}_{ij}({\bf r},t)\hat {G}^<_{ji}(t,t)+i\sum_{iji_1j_1}\int dt''{\bf j}_{0ij}({\bf r}){\bf j}_{0j_1i_1}[\hat {G}_{jj_1}(t,t'')\hat { G}_{i_1i}(t'',t)]^<,
    \]
    where ${\bf j}_{0ij}({\bf r})\equiv \frac 12\lim_{{\bf r}\rightarrow {\bf r}'}[\hat {\bf j}_0({\bf r})+\hat {\bf j}_0^+({\bf r}')]\psi_j({\bf r})\psi^*_i({\bf r}')$ and $\hat {\bf j}_{0ij}=\int d{\bf r}{\bf j}_{0ij}({\bf r})$ is the element of matrix $\hat {\bf j}_0$. $\delta {\bf j}_{ij}({\bf r})$ is defined in the same manner as ${\bf j}_{0ij}({\bf r})$ but with replacing $\hat {\bf j}({\bf r},t)$ and $\hat {\bf j}^+({\bf r}',t)$ by operators $\delta\hat {\bf j}({\bf r})$ and $\delta\hat {\bf j}^+({\bf r}')$, respectively. Performing Fourier transform, the observed current in $({\bf q},\omega)$ space is given by
  \[
    {\bf J}({\bf q},\omega)=-i\sum_{ij}\delta {\bf j}_{ij}({\bf q},\omega)\int \frac{d\omega'}{2\pi}\hat {G}^<_{ji}(\omega)+i\sum_{iji_1j_1}\int \frac{d\omega_1}{2\pi}{\bf j}_{0ij}({\bf q}){\bf j}_{0j_1i_1}[\hat {G}_{jj_1}(\omega_1)\hat { G}_{i_1i}(\omega_1-\omega)]^<.
    \]
Setting ${\bf q}=0$ and using the relation $J_{\alpha}(\omega)=i\omega\sum_{\beta=x,y,z}\sigma_{\alpha\beta}A_\beta(\omega)$ ($\sigma_{\alpha\beta}$ is the conductivity) and Kubo-Martin-Schwinger relation $\hat G^<=n_F(\omega) [\hat G^a(\omega)-\hat G^r(\omega)]$, we finally arrive at
  \begin{align}
    \sigma_{\alpha\beta}(\omega)=&\frac{ie^2}{\omega}\sum_{ij}\frac{\partial^2}{\partial p_\alpha\partial p_\beta}[\hat h_0({\bf p})]_{ij}\int \frac{d\omega'}{2\pi}[-i\hat {G}^<_{ji}(\omega)] +\frac{1}{\omega}\sum_{iji_1j_1}\int \frac{d\omega_1}{2\pi}\hat j^\alpha_{0ij}\hat j^\beta_{0j_1i_1}\{n_F(\omega_1-\omega)[\hat {G}^r_{jj_1}({ G}^a_{i_1i}-{ G}^r_{i_1i})]_{\omega_1,\omega_1-\omega}\nonumber\\
    &+n_F(\omega_1)[({ G}^a_{jj_1}-{ G}^r_{jj_1})\hat {G}^a_{i_1i}]_{\omega_1,\omega_1-\omega}\}.\label{A3}
  \end{align}
In this equation, the first term is just the diamagnetic term. It reduces to $iN_ee^2/(m\omega)$ ($N_e$ is the carrier density) for a one-band Hamiltonian $\hat h({\bf p})={\bf p}^2/(2m)$ but it vanishes when the free-carrier Hamiltonian depends linearly on ${\bf p}$. This implies that the diamagnetic term in conductivity is absent in Dirac-fermion systems, such as the systems with carriers near the Dirac nodes of graphene, silicene, germanene {\it etc.} 

Note that, Eq.\,(\ref{A3}) contains momentum integrations which are implicitly involved in the summations. Due to the specific momentum dependence of $\hat {\bf j}_0$, in Eq.\,(\ref{A3}), only the sums of the terms associated with real parts of quantity $\hat j^\alpha_{0ij}\hat j^\alpha_{0j_1i_1}$ are nonvanishing. Thus, the real part of diagonal conductivity, ${\rm Re}\sigma_{\alpha\alpha}$, can be further rewritten in a compact form:
\begin{align}
  {\rm Re} \sigma_{\alpha\alpha}=&\frac{1}{\omega}\sum_{iji_1j_1}\int \frac{d\omega_1}{2\pi}[n_F(\omega_1)-n_F(\omega_1+\omega)]
  \left \{{\rm Re}(\hat j^\alpha_{0i_1j}\hat j^\alpha_{0j_1i}+\hat j^\alpha_{0ij}\hat j^\alpha_{0j_1i_1}){\rm Im}G_{jj_1}^r(\omega_1+\omega){\rm Im}G^r_{i_1i}(\omega_1)\right .\nonumber\\
  &\left . +{\rm Re}(\hat j^\alpha_{0i_1j}\hat j^\alpha_{0j_1i}-\hat j^\alpha_{0ij}\hat j^\alpha_{0j_1i_1}){\rm Re}G_{jj_1}^r(\omega_1+\omega){\rm Re}G^r_{i_1i}(\omega_1)\right \}.
\end{align}
Here, $\hat G^r_{ij}=(\hat G^a_{ji})^*$ is used.
\end{widetext}

\bibliography{IB}
\end{document}